\documentclass[lettersize,journal]{IEEEtran}
\usepackage{amsmath,amsfonts}
\usepackage{algorithmic}
\usepackage{algorithm}
\usepackage{array}
\usepackage[caption=false,font=normalsize,labelfont=sf,textfont=sf]{subfig}
\usepackage{textcomp}
\usepackage{stfloats}
\usepackage{url}
\usepackage{verbatim}
\usepackage{graphicx}
\usepackage{cite}
\newcommand {\Define} {\stackrel {\Delta} {=}  }

\begin{document}

\title{Zak-OTFS and Turbo Signal Processing for Joint Sensing and Communication}

\author{Jinu Jayachandran, Muhammad Ubadah, Saif Khan Mohammed, Ronny Hadani,
Ananthanarayanan Chockalingam and Robert Calderbank, ~\IEEEmembership{Fellow,~IEEE}
\thanks{J. Jayachandran, M. Ubadah  and S. K. Mohammed are with Department of Electrical Engineering, Indian Institute of Technology Delhi, India (E-mail: Jinu.Jayachandran@ee.iitd.ac.in, Muhammad.Ubadah@ee.iitd.ac.in, saifkmohammed@gmail.com). S. K. Mohammed is also associated with Bharti School of Telecom. Technology and Management (BSTTM), IIT Delhi. R. Hadani is with Department of Mathematics, University of Texas at Austin, TX, USA (E-mail: hadani@math.utexas.edu). A. Chockalingam is with Department of Electrical Communication Engineering, Indian Institute of Science Bangalore, India (E-mail: achockal@iisc.ac.in). R. Calderbank is with Department of Electrical and Computer Engineering, Duke University, USA (E-mail: robert.calderbank@duke.edu).}
\thanks{The work of S. K. Mohammed was supported in part by the Jai Gupta Chair at I.I.T. Delhi, and also in part by a project at BSTTM funded by Bharti Airtel Limited.}
\thanks{This work has been submitted to the IEEE for possible publication. Copyright may be transferred without notice, after which this version may no longer be accessible.}}



\maketitle

\begin{abstract}
The Zak-OTFS input/output (I/O) relation is predictable and non-fading when the delay and Doppler periods are greater than the effective channel delay and Doppler spreads, a condition which we refer to as the crystallization condition. The filter taps can simply be read off from the response to a single Zak-OTFS pilot pulsone, and the I/O relation can be reconstructed for a sampled system that operates under finite duration and bandwidth constraints. In previous work we had measured BER performance of a baseline system where we used separate Zak-OTFS subframes for sensing and data transmission. In this Letter we demonstrate how to use turbo signal processing to match BER performance of this baseline system when we integrate sensing and communication within the same Zak-OTFS subframe. The turbo decoder alternates between channel sensing using a noise-like waveform (spread pulsone) and recovery of data transmitted using point pulsones.
\end{abstract}


\section{Introduction}
6G propagation environments are changing the balance between time-frequency methods focused on OFDM signal processing and delay-Doppler methods (OTFS) \cite{intro1, intro2, mcotfs1, mcotfs2, mcotfs3}. OFDM is configured to prevent inter-carrier-interference (ICI) whereas OTFS is configured to embrace ICI \cite{embed1, embed2, mcpulse1, mcpulse2, derotfs, zak3}. In OFDM, once the I/O relation is known, equalization is relatively simple at least when there is no ICI. However, acquisition of the I/O relation is non-trivial and dependent on the accuracy of the assumed propagation model \cite{ofdm1}. In contrast, equalization is more involved in OTFS due to inter-symbol-interference (ISI), however acquisition of the I/O relation is simple and model free \cite{zakotfs1, zakotfs2, ddpulse1, ddpulse2}. Acquisition becomes more critical when Doppler spreads measured in KHz make it more and more difficult to estimate channels. The most challenging situation is the combination of unresolvable paths and high channel spreads. In this Letter we present simulation results for a Veh-A channel \cite{EVAITU} where the first three paths are not separable and cannot be estimated accurately. 

In previous work \cite{zakotfs1, zakotfs2} we have described how to design a parametric family of pulsone waveforms that can be matched to the delay and Doppler spreads of different propagation environments. A pulsone is a signal in the time domain which realizes a quasi-periodic localized function on the DD domain. The prototypical structure of a pulsone is a train of pulses modulated by a tone. We have analyzed performance in the situation when the pulsone parameters matches the environment channel parameters, in the sense that, the delay period of the pulsone is greater than the delay spread of the channel, and the Doppler period of the pulsone is greater than the Doppler spread of the channel. We refer to this condition as the crystallization condition. We start from this baseline system where we dedicate separate Zak-OTFS subframes to channel estimation and data transmission, abbreviated as $S \vert \overline{C}$ (sensing/channel estimation in the absence of communication signal) and $C \vert \overline{S}$ (communication in the absence of sensing signal).

The characteristic structure of a pulsone is a train of pulses modulated by a tone, a signal with unattractive peak-to-average power ratio (PAPR). In more recent work \cite{zakotfs3} we have introduced the notion of filtering in the discrete delay-Doppler domain. We have described how to construct spread waveforms with desirable characteristics by applying a chirp filter in the discrete DD domain to a point pulsone. One desirable characteristic is low PAPR, about 6dB for the exemplar spread pulsone, compared with about 15dB for the point pulsone. A second desirable characteristic is the ability to read off the I/O relation of the sampled communication system provided a second crystallization condition is satisfied. This work demonstrates how to integrate sensing and communication within a single Zak-OTFS subframe by combining a spread pulsone for channel sensing with point pulsones for data transmission. The filter in the discrete DD domain enables coexistence by minimizing interference between sensing and data transmission. We have demonstrated that sharing DD domain resources in this way increases effective throughput compared with traditional approaches that use guard bands to divide DD domain resources between sensing and communication. 

In this Letter we demonstrate that turbo signal processing is able to close the performance gap between separate sensing and communication ($S \vert \overline{C}$ and $C \vert \overline{S}$) and joint sensing and communication in the same Zak-OTFS subframe (which we abbreviate as $S \vert C$ and $C \vert S$). The turbo principle of iterating between functional blocks in communication receivers has proven to be a powerful method of improving performance. For example, turbo iterations between channel equalizer and channel decoder have been shown to yield tremendous improvements in bit error performance \cite{R1}, \cite{R2}. Likewise, iterations between channel estimation and turbo equalizer have been shown to improve the channel estimate over the iterations by using soft information fed back from the decoder from the previous iteration to generate extended training sequences between the actual transmitted training sequences \cite{R3}.
              \begin{figure}
     \centering
        \includegraphics[width=8.4cm,height=5.5cm]{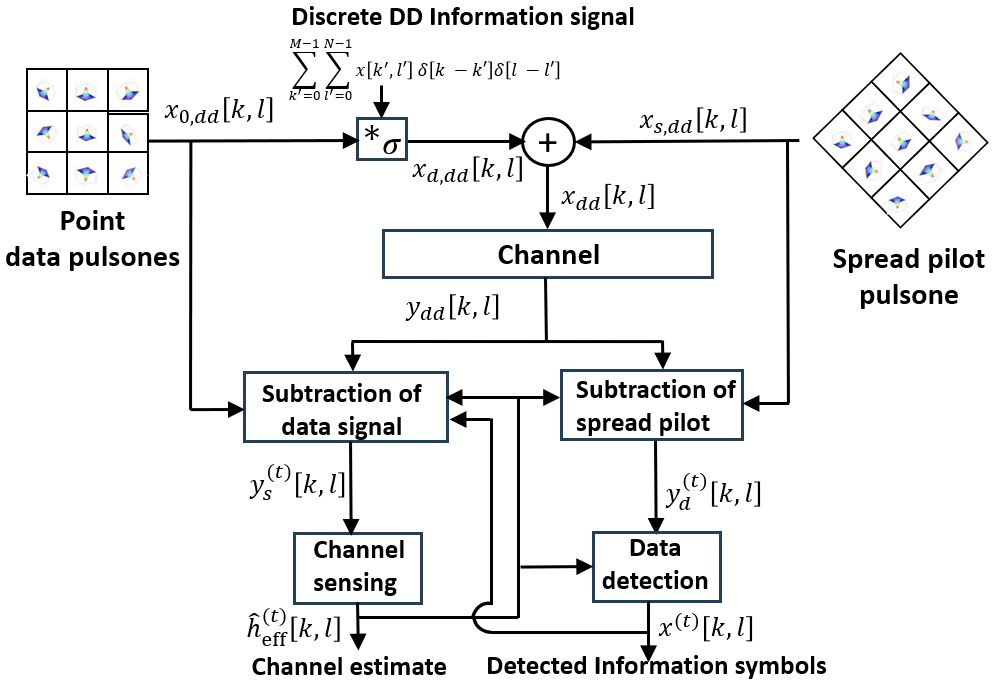}
        \vspace{-2mm}
        \caption{Signal processing for proposed Zak-OTFS based iterative joint sensing and communication.}
        \label{fig_turbo}
        \vspace{-5mm}
    \end{figure} 
In joint sensing and communication, we estimate the effective channel, then estimate the received spread pilot, then recover the data after subtracting our estimate for the received pilot from the received signal (see Fig.~\ref{fig_turbo}). The residual pilot (after cancellation) interferes with data transmission. In the turbo iteration, we take the estimated data, then estimate the received data signal, then improve our estimate for the effective channel by subtracting our estimate for the received data signal from the received signal. We show that five turbo iterations suffice to match the performance of separate sensing and communication across a wide range of Doppler shifts.    
\section{System model}
\label{secsysmodel}
In previous work \cite{zakotfs3} we have introduced a framework for joint sensing and communication where the pilot/sensing signal and the data signal are embedded in a single Zak-OTFS subframe. We recall \cite{zakotfs3} that Zak-OTFS carrier waveforms are quasi-periodic functions in the discrete DD domain with period $M$
along the discrete delay axis and period $N$ along the discrete Doppler axis. The superposition of pilot and data signals in the discrete DD domain is given by
\begin{eqnarray}
\label{eqn30_1}
    x_{\mbox{\scriptsize{dd}}}[k,l] & = & \sqrt{E_d} \, x_{\mbox{\scriptsize{d,dd}}}[k,l] \, + \, \sqrt{E_p} \, x_{\mbox{\scriptsize{s,dd}}}[k,l]
\end{eqnarray}where $E_p$ denotes the energy of the pilot and $E_d$ the energy of the data signal. The ratio $E_p/E_d$ is therefore the pilot to data power ratio (PDR). The data signal $x_{\mbox{\scriptsize{d,dd}}}[k,l]$, $k,l \in {\mathbb Z}$ is given by
\begin{eqnarray}
\label{eqnxddklinf}
    x_{\mbox{\scriptsize{d,dd}}}[k,l] & = & \frac{1}{\sqrt{MN}} \, x[k \, \mbox{\footnotesize{mod}} \, M \, , \, l \, \mbox{\footnotesize{mod}} \, N] \, e^{j2 \pi \frac{\left\lfloor \frac{k}{M} \right\rfloor l }{N} },
\end{eqnarray}where each of the $MN$ information symbols $x[k,l]$, $k=0,\cdots, M-1$, $l=0,1,\cdots, N-1$ have average energy ${\mathbb E}\left[ \vert x[k,l] \vert^2 \right] = 1$. The exponential phase term in (\ref{eqnxddklinf}) renders $x_{\mbox{\scriptsize{d,dd}}}[k,l]$ quasi-periodic. For all $n,m \in {\mathbb Z}$
\begin{eqnarray}
    x_{\mbox{\scriptsize{d,dd}}}[k+nM,l+mN]  & = & e^{j 2 \pi \frac{n l}{N}} \, x_{\mbox{\scriptsize{d,dd}}}[k,l].  
\end{eqnarray}The data signal in (\ref{eqn30_1})
can also be expressed in terms of filtering (in the DD domain) a quasi-periodic DD domain pulse at the origin.
{
\begin{eqnarray}
\label{eqn91654}
    x_{\mbox{\scriptsize{d,dd}}}[k,l] & \hspace{-3mm} = & \hspace{-3mm} {\Bigg [} \, \left( \sum\limits_{k'=0}^{M-1} \sum\limits_{l'=0}^{N-1} x[k',l'] \, \delta[k - k'] \, \delta[l - l'] \right) \, \nonumber \\
    & & \hspace{3mm} *_{\sigma} \, x_{\mbox{\scriptsize{0,dd}}}[k,l] \, {\Bigg ]},
\end{eqnarray}}where
{
\begin{eqnarray}
x_{\mbox{\scriptsize{0,dd}}}[k,l] & \Define & \sum\limits_{n,m \in {\mathbb Z}} \delta[k - nM] \, \delta[l - mN]
\end{eqnarray}}is the data pulsone corresponding to a DD pulse at the origin.
DD domain filtering is implemented through twisted convolution $*_{\sigma}$ and the filter taps in (\ref{eqn91654}) depend on the information symbols.
The pilot signal in (\ref{eqn30_1}) is given by
\begin{eqnarray}
\label{eqn_20}
     x_{\mbox{\scriptsize{s,dd}}}[k,l] & = & w_s[k,l] \, *_{\sigma} \, x_{\mbox{\scriptsize{p,dd}}}[k,l],
\end{eqnarray}where the spreading filter $w_s[k,l]$
acts on the point pilot $x_{\mbox{\scriptsize{p,dd}}}[k,l] $ by twisted convolution. We follow \cite{zakotfs3} by considering chirp filters, where for $k=0,1,\cdots, M-1$, $l=0,1,\cdots, N-1$ 
\begin{eqnarray}
\label{eqnwpchirp}
    w_s[k,l] & = &  \frac{1}{MN} \, e^{j 2 \pi \frac{q (k^2 + l^2)}{MN}}, 
\end{eqnarray}and we refer to $q \in {\mathbb Z}$ as the slope parameter. The point pilot $x_{\mbox{\scriptsize{p,dd}}}[k, l]$ appearing in (\ref{eqn_20}) is a discrete DD domain quasi-periodic impulse located at $(k_p, l_p)$ and is given by
\begin{eqnarray}
\label{xpddprobeeqn}
x_{\mbox{\scriptsize{p,dd}}}[k, l] & \hspace{-3mm} = & \hspace{-4.5mm}  \sum\limits_{n,m \in {\mathbb Z}} \hspace{-2mm} e^{j 2 \pi \frac{n l_p}{N}} \, \delta[k - k_p - nM] \, \delta[l - l_p - mN].
\end{eqnarray}

We then \emph{lift} the discrete DD domain signal
$x_{\mbox{\scriptsize{dd}}}[k, l]$ given by (\ref{eqn30_1}) to obtain a \emph{continuous} DD domain
signal
\begin{eqnarray}
    x_{\mbox{\scriptsize{dd}}}(\tau, \nu) = \sum\limits_{k,l \in {\mathbb Z}} x_{\mbox{\scriptsize{dd}}}[k, l] \delta(\tau - k \tau_p/M) \delta(\nu - l \nu_p/N)
\end{eqnarray}that is quasi-periodic with delay period $\tau_p$ and Doppler peirod $\nu_p = 1/\tau_p$.
We then apply a pulse shaping filter $w_{tx}(\tau, \nu)$
to limit the transmitted TD signal to the duration and bandwidth of the Zak-OTFS subframe. This TD signal is obtained by applying the inverse Zak transform to the filtered DD domain signal $w_{tx}(\tau, \nu) \, *_{\sigma} \, x_{\mbox{\scriptsize{dd}}}(\tau, \nu) $. 

Matched filtering at the receiver using
$w_{rx}(\tau, \nu) = w_{tx}^*(-\tau, -\nu) \, e^{j 2 \pi \nu \tau}$ followed by sampling on the information grid results in the discrete DD domain signal given by
{
\begin{eqnarray}
\label{eqn87254}
y_{\mbox{\scriptsize{dd}}}[k, l] 
& \hspace{-2mm}  =  &   \hspace{-2mm} \underbrace{\sqrt{E_d} \, h_{\mbox{\scriptsize{eff}}}[k, l] *_{\sigma} x_{\mbox{\scriptsize{d,dd}}}[k, l]}_{\mbox{\small{Rx. data signal}}}  \nonumber \\
& & \hspace{-1mm}  +   \underbrace{\sqrt{E_p} h_{\mbox{\scriptsize{eff}}}[k, l] *_{\sigma} x_{\mbox{\scriptsize{s,dd}}}[k, l]}_{\mbox{\small{Rx. sensing/pilot signal}}} \,  + \, n_{\mbox{\scriptsize{dd}}}[k, l],
\end{eqnarray}}where $n_{\mbox{\scriptsize{dd}}}[k, l]$
represents noise in the discrete DD domain. Note that it is associativity of twisted convolution that allows us to represent a cascade of filters/channels as a single effective channel filter $h_{\mbox{\scriptsize{eff}}}[k, l]$.
In our previous work \cite{zakotfs3} we have described how the receiver senses the channel (estimates $h_{\mbox{\scriptsize{eff}}}[k, l]$) from the cross-ambiguity $A_{y, x_s}[k,l]$ between $y_{\mbox{\scriptsize{dd}}}[k, l]$ and the transmitted spread pilot $x_{\mbox{\scriptsize{s,dd}}}[k,l]$.
We recall (Theorem $3$ from \cite{zakotfs3}) that
{
\begin{eqnarray}
\label{eqn352_ays}
    { A}_{y,x_s}[k,l]  & \hspace{-2mm} = &  \hspace{-2mm}  \sqrt{E_p} \, h_{\mbox{\scriptsize{eff}}}[k,l] \, *_{\sigma} \, { A}_{x_s,x_s}[k,l] \nonumber \\
    & &  \hspace{-3mm} +  \sqrt{E_d} \,  h_{\mbox{\scriptsize{eff}}}[k,l] \, *_{\sigma} \, {A}_{x_d,x_s}[k,l] \, + \, { A}_{n,x_s}[k,l],\nonumber \\
\end{eqnarray}}where ${ A}_{x_s,x_s}[k,l]$ is the self-ambiguity function of the spread pilot signal, ${A}_{x_d,x_s}[k,l]$ and ${A}_{n,x_s}[k,l]$ are the cross-ambiguity functions of the transmitted data signal $x_{\mbox{\scriptsize{d,dd}}}[k,l]$ and noise signal $n_{\mbox{\scriptsize{dd}}}[k,l]$ respectively with the spread pilot signal.

The self-ambiguity function of the point pilot  $x_{\mbox{\scriptsize{p,dd}}}[k,l]$
is supported on the \emph{period lattice} $\Lambda_p$ comprising integer linear combinations of $(\tau_p, 0)$ and $(0, \nu_p)$.

We have shown \cite{zakotfs3} that the self-ambiguity function of the spread pilot is supported on a lattice $\Lambda$ obtained by rotating $\Lambda_p$. Let ${\mathcal S}$ denote the support region of the effective channel in the discrete DD domain. The crystallization conditions with respect to the lattice $\Lambda$ are satisfied if the translates of ${\mathcal S}$ by lattice points in $\Lambda$ are disjoint. In this case, we can obtain the effective channel tap $h_{\mbox{\scriptsize{eff}}}[k,l]$ by evaluating the first term of (\ref{eqn352_ays}) at points $(k,l)$ in a fundamental domain of $\Lambda$. The second term in (\ref{eqn352_ays}) represents interference to sensing from data. 
We recall (Theorem $4$ from \cite{zakotfs3}) that the magnitudes $\vert { A}_{x_d,x_s}[k,l] \vert$ are essentially independent of $k,l$ so that interference from data to sensing is noise-like (see also Fig.$18$
from \cite{zakotfs3}). The third term in (\ref{eqn352_ays}) represents interference to sensing from noise. We recall (Appendix J from \cite{zakotfs3}) that ${ A}_{n,x_s}[k,l]$ is zero-mean Gaussian distributed with variance essentially independent of $k,l$. Our estimate of $h_{\mbox{\scriptsize{eff}}}[k,l]$ is then
\begin{eqnarray}
\label{eqnest23}
    {\widehat h}_{\mbox{\scriptsize{eff}}}[k,l] & = & \frac{{ A}_{y,x_s}[k,l]}{\sqrt{E_p}} \,\,\, \mbox{\small{for}} \,\, k,l \in {\mathcal S}.
\end{eqnarray}We suppose that the crystallization conditions are also satisfied 
with respect to the lattice $\Lambda_p$. We use the estimate (\ref{eqnest23}) to cancel the contribution made by the pilot to the received DD signal. After cancellation, the signal
\begin{eqnarray}
\label{pilotfreeeqn}
     y_{\mbox{\scriptsize{dd}}}[k,l] -  \sqrt{E_p} \, {\widehat h}_{\mbox{\scriptsize{eff}}}[k,l] *_{\sigma} x_{\mbox{\scriptsize{s,dd}}}[k,l].
\end{eqnarray}is used to recover the data using the matrix-vector formulation of the Zak-OTFS I/O relation (see \cite{zakotfs2} for details).
\begin{figure*}
\begin{eqnarray}
\label{eqn196402}
    {\widehat y}_{\mbox{\scriptsize{d,dd}}}^{(t)}[k,l] & = &  {\widehat h}_{\mbox{\scriptsize{eff}}}^{(t-1)}[k,l] *_{\sigma} \left[  \sum\limits_{k'=0}^{M-1} \sum\limits_{l'=0}^{N-1} {\widehat x}^{(t-1)}[k',l'] \delta[k - k'] \, \delta[l - l']  *_{\sigma}  {x}_{\mbox{\scriptsize{0,dd}}}[k,l] \right].
\end{eqnarray}
\begin{eqnarray*}
\hline
\end{eqnarray*}
\end{figure*}
By spreading the pilot signal, we integrate sensing and communication within the same Zak-OTFS subframe. Sharing DD domain resources in this way increases effective throughput compared with traditional approaches that use
guard bands to divide DD domain resources between sensing and communication (see Fig.$28$ from \cite{zakotfs3}). Spreading also reduces the PAPR of the pilot signal to about $5$ dB compared with $15$ dB for the original point pilot (see Fig.$10$ and Fig.$11$ from \cite{zakotfs3}).

We recall that integrated sensing and communication with spread pilots results in an uncoded $4$-QAM BER of about $10^{-2}$,
compared with $10^{-5}$ for sensing and communication in separate Zak-OTFS subframes (see Fig.$26$ from \cite{zakotfs3}). The difference is three orders of magnitude. Section \ref{sec3p3} describes a turbo signal processing method that is able to close this gap.

\section{Iterative Sensing and Communication}
\label{sec3p3}
In this Section we describe the iterative signal processing algorithm illustrated in Fig.~\ref{fig_turbo}. The first iteration is described in Section \ref{secsysmodel} and the $t^{th}$ iteration consists of four steps. 

\textbf{STEP $1$}: Inputs are the detected information symbols ${\widehat x}^{(t-1)}[k,l]$, $k=0,1,\cdots, M-1$, $l=0,1,\cdots, N-1$ and estimated channel filter taps
${\widehat h}_{\mbox{\scriptsize{eff}}}^{(t-1)}[k,l], (k,l) \in {\mathcal S}$ from
iteration $t-1$. We form ${\widehat y}_{\mbox{\scriptsize{d,dd}}}^{(t)}[k,l]$ (see (\ref{eqn196402}) at top of page)
and subtract this estimate for the received data signal from the received signal to obtain
\begin{eqnarray}
    y_s^{(t)}[k,l] & = & y_{\mbox{\scriptsize{dd}}}[k,l] \, - \, {\widehat y}_{\mbox{\scriptsize{d,dd}}}^{(t)}[k,l].
\end{eqnarray}The output $y_s^{(t)}[k,l]$
is a \emph{data cancelled signal} that approximates the received spread pilot. 

\textbf{STEP $2$}: We use (\ref{eqnest23})
to form the $t^{th}$ estimate ${\widehat h}_{\mbox{\scriptsize{eff}}}^{(t)}[k,l]$ for the effective channel filter taps
\begin{eqnarray}
    {\widehat h}_{\mbox{\scriptsize{eff}}}^{(t)}[k,l] & = & \frac{A_{y_s^{(t)}, x_s}[k,l]}{\sqrt{E_p}} \,\, \mbox{\small{for}} \,\, (k,l) \in {\mathcal S}.
\end{eqnarray}

\textbf{STEP $3$}: We use ${\widehat h}_{\mbox{\scriptsize{eff}}}^{(t)}[k,l]$, $(k,l) \in {\mathbb S}$ to form the $t^{th}$ estimate ${\widehat y}_{\mbox{\scriptsize{s,dd}}}^{(t)}[k,l]$ of the received pilot signal using
\begin{eqnarray}
{\widehat y}_{\mbox{\scriptsize{s,dd}}}^{(t)}[k,l] & = & {\widehat h}_{\mbox{\scriptsize{eff}}}^{(t)}[k,l] \, *_{\sigma} \, {x}_{\mbox{\scriptsize{s,dd}}}[k,l].
\end{eqnarray}We subtract this estimate from the received signal to obtain
\begin{eqnarray}
    y_d^{(t)}[k,l] & = & y_{\mbox{\scriptsize{dd}}}[k,l] \, - \, {\widehat y}_{\mbox{\scriptsize{s,dd}}}^{(t)}[k,l].
\end{eqnarray}This output is a \emph{pilot cancelled signal} that approximates the received data signal. 

\textbf{STEP $4$}: We detect data/information symbols $\widehat{x}^{(t)}[k,l]$ from $y_d^{(t)}[k,l]$ using the method described in Section \ref{secsysmodel}. We then move to STEP $1$ of iteration $t+1$, and the algorithm halts after a fixed number of iterations. 

Section \ref{simsec} presents numerical simulations illustrating that multiple iterations improve the fidelity of the \emph{data cancelled} and \emph{pilot cancelled} signals.

      \begin{table}[]
            \centering
            \caption{Power-delay profile of Veh-A channel model}
            \begin{tabular}{|c|c|c|c|c|c|c|}
                \hline
                Path number ($i$)       &  1 &  2  &  3  &  4  &  5  &  6  \\
                \hline
                $\tau_i$ ($\mu$s) &  0 & 0.31& 0.71& 1.09 & 1.73 & 2.51 \\
                \hline
                Relative power ($p_i$) dB   &  0 & -1  & -9  & -10 & -15 & -20\\
                \hline
            \end{tabular}
            \label{tab_veha}
            \vspace{2mm}
            
            \vspace{-4mm}
        \end{table}
\section{Numerical simulations}
\label{simsec}
This Section reports simulation results for the Veh-A channel model \cite{EVAITU} which consists of six channel paths. The channel gains $h_i$ are modeled as independent zero-mean complex circularly symmetric Gaussian random variables, normalized so that $\sum\limits_{i=1}^6{\mathbb E}\left[ \vert h_i \vert^2 \right] = 1$. Table~\ref{tab_veha} lists the power-delay profile for the six channel paths. The Doppler shift of the $i$-th path is modeled as $\nu_i = \nu_{max} \cos(\theta_i)$, where $\nu_{max}$ is the maximum Doppler shift of any path, and the variables $\theta_i$, $i=1,2, \cdots, 6$, are independent and distributed uniformly in $[-\pi \,,\, \pi)$. 

We consider Zak-OTFS modulation with Doppler period $\nu_p = 30$ KHz, delay period $\tau_p = 1/ \nu_p = 33.3$ ms, $M=31$ and $N=37$. The channel bandwidth $B = M \nu_p = 0.93$ MHz and the subframe duration $T = N \tau_p = 1.2$ ms. Note that the first three channel paths introduce delay shifts in the interval   $[0, 0.71]$ ms and each is less than the delay resolution. We consider matched filtering using root raised cosine (RRC) pulse shaping filters with roll-off factors $\beta_{\nu} = \beta_{\tau} = 0.6$. This increases the subframe duration to $T' = (1 + \beta_{\nu}) T = 1.92$ ms and the bandwidth to $B' = (1 + \beta_{\tau}) B = 1.40$ MHz. We consider a spread pilot pulsone constructed using a chirp filter in the discrete delay-Doppler domain ((\ref{eqnwpchirp}) with $q=3$) as described in \cite{zakotfs3}. We implement the turbo signal processing pipeline illustrated in Fig.~\ref{fig_turbo} using MMSE equalization to recover information symbols using the effective channel matrix.

\begin{figure}
     \centering
        \includegraphics[width=8.7cm,height=5.8cm]{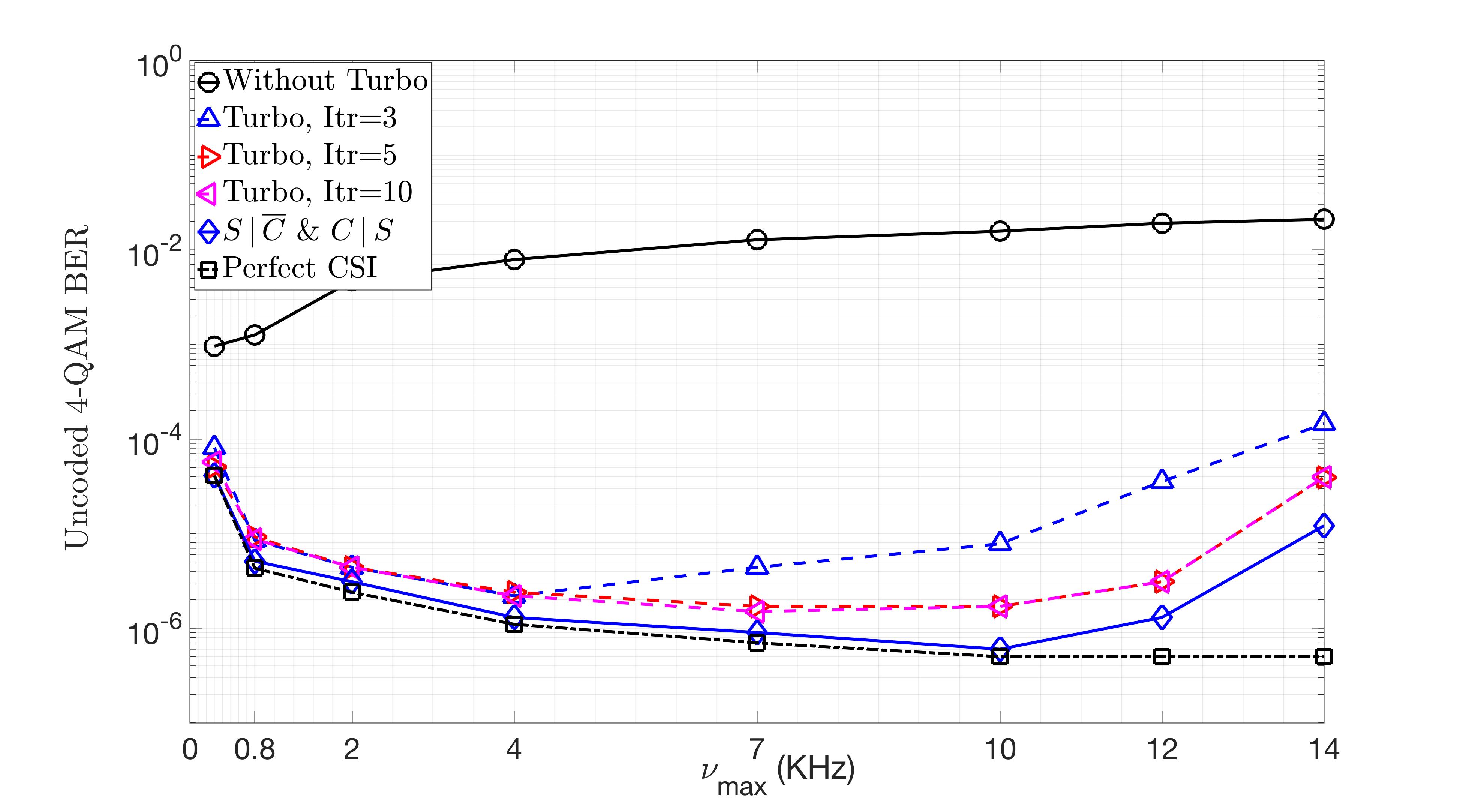}
        \vspace{-2mm}
        \caption{Uncoded $4$-QAM BER as a function of increasing $\nu_{max}$. Veh-A channel, data SNR $=25$ dB, PDR $= 10$ dB, Doppler period $\nu_p = 30$ KHz, $M=31, N=37$, RRC pulse shaping filter                     ($\beta_{\nu} = \beta_{\tau} = 0.6$).}
        \label{fig_bernumax}
        \vspace{-5mm}
    \end{figure}

For a fixed data SNR of $25$ dB we set the PDR to $10$ dB so that the pilot power is sufficient to start the turbo process. Fig.~\ref{fig_bernumax} illustrates that in the crystalline regime, five turbo iterations (dashed curve with red triangles) are sufficient to match the BER performance of separate sensing and communications (dashed curve with black squares). When $\nu_{max} > 12$ KHz the estimate of the effective channel becomes less accurate because of Doppler domain aliasing, and BER performance degrades because interference from the residual pilot becomes more significant than noise. We focus on the role of channel estimation by designing a reference system (solid blue curve with diamonds) where sensing takes place in a separate subframe ($S \vert \overline{C}$) but data transmission is still subject to interference from the residual pilot ($C \vert S$). Fig.~\ref{fig_bernumax} illustrates that BER performance with five turbo iterations differs from that of the reference system by a small SNR offset.

\begin{figure}
     \centering
        \includegraphics[width=8.7cm,height=5.8cm]{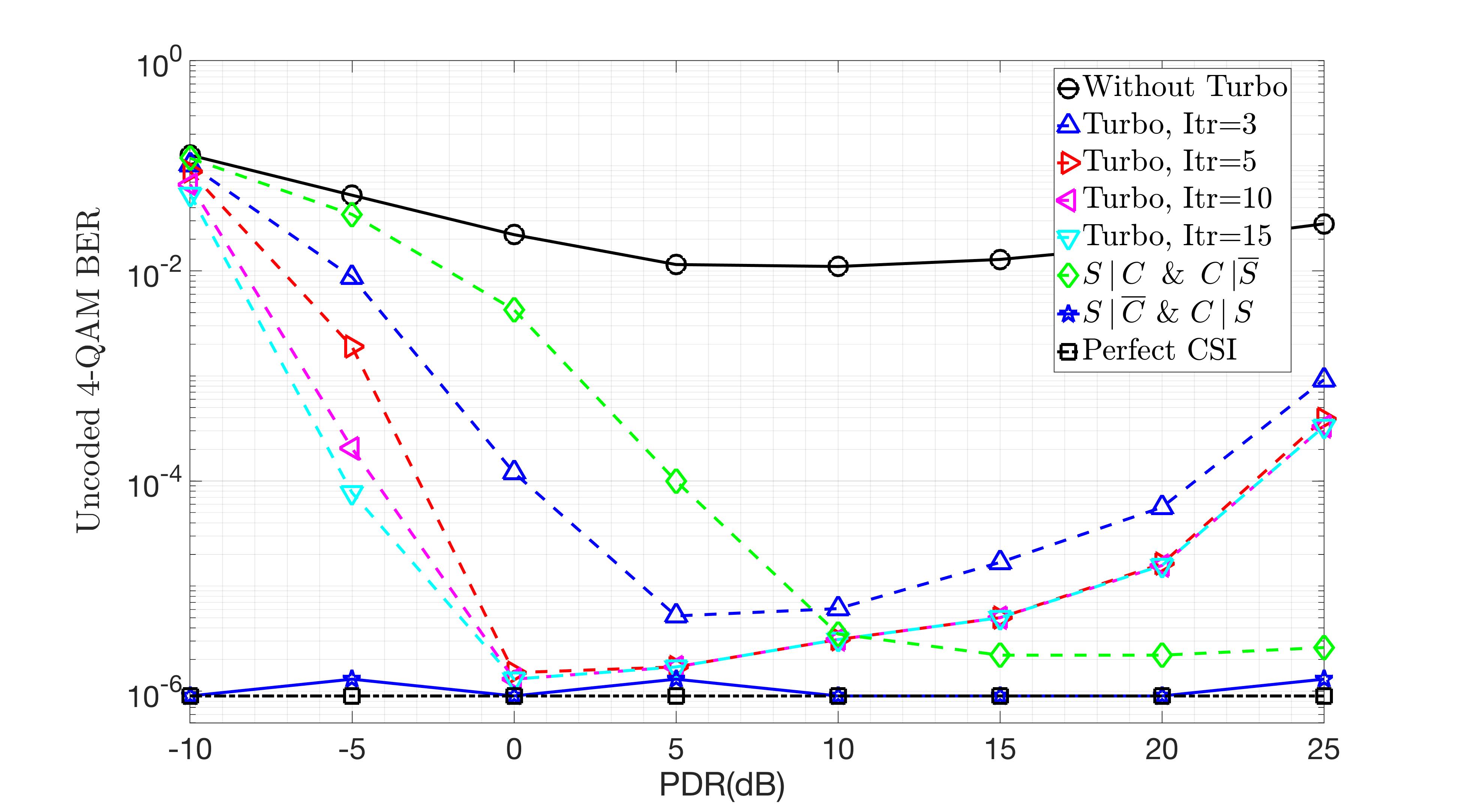}
        \vspace{-2mm}
        \caption{Uncoded $4$-QAM BER as a function of increasing PDR. Veh-A channel considered in Fig.~\ref{fig_bernumax}.}
        \label{fig_bervspdr}
        \vspace{-5mm}
    \end{figure}

 Next we consider how BER performance depends on PDR, and we set $\nu_{max} = 6$ KHz so that we are operating deep in the interior of the crystalline regime. Fig.~\ref{fig_bervspdr} illustrates that when PDR $< -10$ dB the pilot power is not sufficient to start the turbo process, that the initial estimate of the effective channel is not sufficiently accurate. When PDR $> 25$ dB interference from the residual pilot (after cancellation) is more significant than noise, and BER degrades as PDR increases. For intermediate values of the PDR, BER performance improves with increasing PDR as estimates of the effective channel become more accurate. This explains the characteristic “U” shape of the intermediate curves in Fig.~\ref{fig_bervspdr}. Again, we focus on the role of channel estimation by considering the reference system described above ($S \vert \overline{C}$ and $C \vert S$), where the residual pilot has very little energy and interference offered to the transmitted data is inconsequential (channel estimation is very accurate since it is based on a separate sensing-only subframe). We also consider a second reference system (green curve with diamonds), where sensing is subject to interference from data ($S \vert {C}$), but data recovery is not subject to interference from a residual pilot ($C \vert \overline{S}$). Data is transmitted in a separate subframe and there are no turbo iterations for this second reference system.     Fig.~\ref{fig_bervspdr} confirms that interference from the residual pilot is responsible for the degradation in BER performance with increasing PDR. 
 
    \section{Conclusions}
We started by observing that the uncoded 4-QAM BER performance of sensing and communications in separate Zak-OTFS subframes is three orders of magnitude better than that of integrated sensing and communication with spread pilots. In joint sensing and communication, we estimate the effective channel, then estimate the received spread pilot, then recover the data after subtracting our estimate for the received pilot from the received signal. The residual pilot (after cancellation) interferes with data transmission, and we showed that this is the reason for the three orders of magnitude gap in performance. We described a turbo signal processing algorithm that alternates between channel sensing using a spread pilot and data recovery. We showed that five turbo iterations suffice to match the performance of separate sensing and communication across a broad range of Doppler shifts.



\end{document}